\documentclass[aps,pra,reprint,superscriptaddress]{revtex4-2}

\usepackage{amsmath,amsfonts}
\usepackage{xcolor,graphicx}        
\usepackage{braket}
\usepackage{float}
\usepackage{tikz}
\usepackage{hyperref}
\hypersetup{
	colorlinks = true,
	linkcolor = cyan,
	anchorcolor = cyan,
	citecolor = red,
	filecolor = cyan,
	urlcolor = cyan}

\begin{document}

\title{Unitary transformation approach to the paraxial wave equation}

\author{M. Huerta-Sandoval}
\email[e-mail:\,]{montserrat.huerta@inaoe.mx}
\affiliation{Instituto Nacional de Astrofísica Óptica y Electrónica (INAOE)\\ Luis Enrique Erro 1, Santa María Tonantzintla, Puebla, 72840, Mexico}

\author{K. Uriostegui}
\affiliation{Instituto de Ciencias Físicas, Universidad Nacional Autónoma de México, Apartado Postal 48-3, Cuernavaca, Morelos 62251, Mexico}

\author{I. Ramos-Prieto}
\email[e-mail:\,]{iran@inaoep.mx}
\affiliation{Instituto Nacional de Astrofísica Óptica y Electrónica (INAOE)\\ Luis Enrique Erro 1, Santa María Tonantzintla, Puebla, 72840, Mexico}

\author{F. Soto-Eguibar}
\email[e-mail:\,]{feguibar@inaoep.mx}
\affiliation{Instituto Nacional de Astrofísica Óptica y Electrónica (INAOE)\\ Luis Enrique Erro 1, Santa María Tonantzintla, Puebla, 72840, Mexico}

\author{H. Moya-Cessa}
\affiliation{Instituto Nacional de Astrofísica Óptica y Electrónica (INAOE)\\ Luis Enrique Erro 1, Santa María Tonantzintla, Puebla, 72840, Mexico}

\begin{abstract}
We present a framework for the paraxial wave equation based on propagation-dependent unitary transformations closely related to the Lewis-Ermakov invariant. This approach establishes a formal equivalence between free-space propagation and the dynamics in a quadratic gradient index (GRIN) medium. In this context, the dynamical invariant and the free-space Hamiltonian do not commute at the initial propagation stage due to Gaussian modulation, which imposes an effective quadratic confinement. Exact commutativity would only be possible for an infinitely wide, nonsquare-integrable optical field; therefore, any finite-energy beam propagates as if it were subject to a quadratic GRIN-like potential. The unitary transformation approach reveals how the Gaussian envelope of physical beams leads to effective harmonic confinement and connects the propagation dynamics to oscillator-like invariants. This method enables the derivation of stationary solutions in different coordinate systems by mapping to an effective quadratic-like medium and establishes a direct link to the zero-frequency Ermakov equation and the Lewis-Ermakov invariants.
\end{abstract}

\date{\today}

\maketitle
\section{Introduction}
Separation of variables in different coordinate systems is a standard technique for solving the paraxial wave equation and identifying its eigenmodes \cite{Saleh_1991,Laser_Beam_Propagation}. Although the resulting solutions often resemble those of quantized harmonic oscillators, this analogy is not exact because of propagation-dependent effects. For instance, Hermite-Gauss modes in rectangular coordinates correspond to two uncoupled harmonic oscillators, but their beam width evolves during propagation as $W(z) = W_0 \sqrt{1 + z^2 / z_R^2}$ \cite{Boyd_HGB_1961}, in contrast to the fixed width of stationary quantum oscillators. Similar scaling behavior is observed for the Laguerre-Gauss modes in cylindrical coordinates \cite{Allen_LGB_1992,Plick_2015} and for the Ince-Gauss modes in elliptical coordinates \cite{Mardoyan1985,Bandres_Ince_2004}. Although these mode families provide separable solutions to the paraxial wave equation, they do not correspond directly to the stationary states of decoupled harmonic oscillator Hamiltonians, as their profiles explicitly depend on the propagation distance or, equivalently, on time evolution in the paraxial or Schrödinger equation formalism.

When the Hamiltonian contains time-dependent parameters, its eigenvalues and eigenfunctions acquire an explicit dependence on the propagation variable, becoming instantaneous quantities. In quantum mechanics, a systematic framework based on dynamical invariants and unitary transformations has been developed to address such systems, allowing factorization of time dependence and the construction of solutions that evolve according to these invariants \cite{Lewis_1967,Lewis_1969,Leach_1977a,Leach_1977b,Markov,Dodonov_1987,Dodonov_2000,Dodonov_2000b,Ramos_2018b,Urzua_2019,Ramos_2023a}. This connection forms the foundation of the present work, which adapts these quantum mechanical methods to the context of paraxial optics. In this way, the underlying oscillator structure governing the propagation of structured light fields is made explicit. It is also important to note that, across all coordinate systems, the fundamental Gaussian mode plays a central role, as it forms the basis for generating a complete set of structured optical modes through appropriate transformations.

Moreover, the paraxial wave equation in free space is mathematically equivalent to the Schrödinger equation for a free particle. This equivalence enables the application of quantum mechanical methods to beam propagation, particularly unitary transformations developed in quantum optics \cite{Nazarathy:80,Stoler_1981,Nienhuis_1993}. Despite this connection, operator-based approaches; such as those involving commutation relations and Lie algebras, remain uncommon in paraxial optics \cite{Moshinsky_1973,Wolf_1974,Rossmann2002,Wolf_2004,Hall2013,Hall2015,Korneev_2025}. When a transverse quadratic GRIN medium is present, the paraxial equation maps onto a two-dimensional harmonic oscillator, and its solutions can be expressed using the fractional Fourier transform \cite{Ramos_GRIN_2024,Urzua_2024}. More generally, this correspondence allows paraxial optics to be interpreted as a quantum evolution, where the propagation distance plays the role of time, and mode transformations correspond to time-dependent unitary operations. This perspective provides a systematic approach to describe structured light using dynamical invariants and Ermakov-type equations \cite{Lewis_1967,Lewis_1969,Leach_1977a,Leach_1977b,Goncharenko,Bertin}.

At the core of our approach is the use of unitary operators that transform the paraxial wave equation into an equivalent system describing two decoupled quantum harmonic oscillators. Specifically, we employ unitary operators \cite{Nazarathy:80,Stoler_1981,Nienhuis_1993} that map the paraxial wave equation, in rectangular coordinates, onto a system of two independent quantum harmonic oscillators, resulting in a conserved Lewis-Ermakov invariant \cite{Leach_1977a,Leach_1977b,Dodonov_2000,Guasti_2003,Ramos_2023a} despite the explicit propagation dependence of the Hamiltonian. The invariant and the free-space Hamiltonian do not commute at the initial stage due to the quadratic confinement induced by the Gaussian envelope \cite{Boyd_HGB_1961}, which prevents the exact coincidence required for conventional shortcuts to adiabaticity protocols \cite{Leach_1977a,Leach_1977b,Guery_2019}. The Gaussian envelope naturally reconciles free-space propagation with oscillator-like invariants through effective harmonic trapping. This formalism also demonstrates, at the mathematical level, how additional Hamiltonian terms could suppress mode transitions via counter-adiabatic contributions \cite{Berry_2009,Martinez_2017,Guery_2019}, although such modifications would fundamentally change the propagation from free space to mediated dynamics, enabling new beam control strategies within physical constraints.

This work is organized as follows. In Section~\ref{GA}, we establish a general framework based on propagation-dependent unitary transformations, yielding a paraxial equation formally equivalent to a Schrödinger-like equation in a quadratic GRIN-like medium. This formulation provides a unified description of structured light propagation and maps directly onto the dynamics of two uncoupled harmonic oscillators. Within this framework, we analyze three families of coordinate systems that naturally sustain structured beam solutions: Hermite-Gauss, Laguerre-Gauss, and Ince-Gauss modes, each discussed in dedicated subsections. In Section~\ref{LEI}, we examine the role of the Lewis-Ermakov invariant and its connection to beam propagation. The section explores the non-commutativity between the invariant and the free space Hamiltonian at the initial stage, the physical implications of the effective quadratic confinement induced by the Gaussian envelope, and the structure of instantaneous eigenvalues and eigenfunctions. Finally, concluding remarks are presented in Section~\ref{Conclusiones}.
\section{General approach}\label{GA}
In this section, we employ a method based on propagation-dependent unitary transformations, where the propagation distance in the paraxial regime plays the role of time in the Schrödinger equation \cite{Sakurai_2017,Laser_Beam_Propagation}. We begin our analysis in rectangular coordinates, which serve as a natural reference and facilitate the extension of the method to other coordinate systems discussed later. In this representation, the paraxial wave equation that describes the free-space evolution of the optical field $E\left(\mathbf{r}_{\perp},\tau\right)$ is given by 
\begin{equation}\label{Eq_1}
    \begin{split}
    i\frac{\partial}{\partial \tau}E\left(\mathbf{r}_{\perp},\tau\right) &= 
    \frac{1}{2}\left(\hat{p}_x^2 +\hat{p}_y^2\right)
    E\left(\mathbf{r}_{\perp},\tau\right),\\
    &= \hat{H}_{\texttt{free}}\, E\left(\mathbf{r}_{\perp},\tau\right),
    \end{split}
\end{equation}  
where $\mathbf{r}_{\perp} = (x,y)$ are the transverse spatial coordinates and $\tau = z/k$ is the normalized propagation distance, with $k = 2\pi/\lambda$ the wavenumber, and $\hat{H}_{\texttt{free}} = \left(\hat{p}_x^2 +\hat{p}_y^2\right)/2$. The momentum-like operators are defined as $\hat{p}_\mu = -i\partial/\partial \mu$ (for $\mu = x, y$) and satisfy the canonical commutation relations $[\hat{p}_\mu, \mu] = -i$. To recast Eq.~\eqref{Eq_1} in a form equivalent to two decoupled quantum harmonic oscillators, we introduce two unitary transformations, both dependent on $\tau$:  
\begin{equation}\label{Transformations_GS}
\begin{split}
\hat{G}(\tau) & = \exp\left(i\frac{\dot{\rho}}{2\rho}\left(x^2 + y^2\right)\right), \\
\hat{S}(\tau) & = \exp\left(-i\frac{\ln(\rho)}{2}\left(x\hat{p}_x + \hat{p}_x x + y\hat{p}_y + \hat{p}_y y\right)\right),
\end{split}
\end{equation}
where $\rho = \rho(\tau)$ is a real-time-dependent scaling function and $\dot{\rho} = d\rho/d\tau$. These operators implement successive frame transformations such that $E(\mathbf{r}_{\perp}, \tau) = \hat{G}(\tau) E_G(\mathbf{r}_{\perp}, \tau)$ and $E_G(\mathbf{r}_{\perp}, \tau) = \hat{S}(\tau) E_S(\mathbf{r}_{\perp}, \tau)$, with $E_G$ and $E_S$ denoting the field in intermediate and final transformed frames, respectively. The structure of these transformations is closely related to the formalism of quantum harmonic oscillators with time-dependent parameters and corresponds to the zero-frequency limit of the Lewis–Ermakov invariant \cite{Bagchi_2021,Bertin}, thus enabling a reinterpretation of paraxial beam propagation in terms of dynamical invariants in quantum mechanics.

Therefore, the successive application of unitary transformations to Eq.~\eqref{Eq_1} yields the paraxial wave equation in a doubly transformed frame, which at each step takes the form:
$i\frac{\partial}{\partial \tau}\psi = (\hat{T}^\dagger \hat{H} \hat{T} - i \hat{T}^\dagger \frac{\partial\hat{T}}{\partial\tau})\psi$, where $\hat{T}=\hat{T}(\tau)$ is a propagation-dependent unitary operator, while $\hat{H}$ and $\psi$ denote the original Hamiltonian and the transformed field, respectively. In the first step, the transformation $E(\mathbf{r}_\perp,\tau) = \hat{G}(\tau)E_G(\mathbf{r}_\perp,\tau)$ leads to
\begin{equation}
\begin{split}
\hat{H}_G(\tau) &= \hat{G}^\dagger \hat{H}_{\texttt{free}} \hat{G} - i \hat{G}^\dagger \frac{\partial\hat{G}}{\partial \tau},\\
&=\frac{1}{2} \sum_{\mu=x,y}\left[\hat{p}_\mu^2 + \frac{\ddot{\rho}}{\rho}\mu^2+\frac{\dot{\rho}}{\rho}\left(\mu\hat{p}_\mu + \hat{p}_\mu \mu \right)\right],
\end{split}
\end{equation}
with $\hat{G}^\dagger \hat{p}_\mu \hat{G} = \hat{p}_\mu + (\dot{\rho}/\rho)\mu$. Subsequently, the second transformation $E_G(\mathbf{r}_\perp,\tau) = \hat{S}(\tau)E_S(\mathbf{r}_\perp,\tau)$ yields
\begin{equation}
\begin{split}
\hat{H}_S(\tau) &= \hat{S}^\dagger \hat{H}_{G}(\tau)\, \hat{S} - i \hat{S}^\dagger \frac{\partial \hat{S}}{\partial \tau}=\frac{1}{2}\sum_{\mu=x,y}\left(\frac{\hat{p}_\mu^2}{\rho^2} + \ddot{\rho}\rho\mu^2\right),
\end{split}
\end{equation}
using the relations $\hat{S}^\dagger \mu \hat{S} = \rho \mu$ and $\hat{S}^\dagger \hat{p}_\mu \hat{S} = \hat{p}_\mu/\rho$, with $\mu = x,y$, which directly leads to the evolution equation for $E_S(\mathbf{r}_\perp,\tau)$,
\begin{equation}\label{H_S}
    i\frac{\partial}{\partial\tau}E_S(\mathbf{r}_\perp,\tau) = \frac{1}{2}\left[\frac{(\hat{p}_x^2 + \hat{p}_y^2)}{\rho^2} + \ddot{\rho}\rho(x^2 + y^2)\right]E_{S}(\mathbf{r}_\perp,\tau).
\end{equation}
These transformations connect the original Hamiltonian to that of a time-dependent harmonic oscillator, where the auxiliary function $\rho(\tau)$ satisfies the Ermakov equation $\ddot{\rho} + \omega^2(\tau)\rho = b_0^2/\rho^3$ \cite{Bertin,Ramos_2023a}. In optical systems with a parabolic refractive index profile, this relation naturally arises from the complex Riccati equation, which can be recast as the real Ermakov equation \cite{Cruz_HGB_2017}. In the present case, Eq.~\eqref{Eq_1} corresponds to free-space evolution (a free particle), implying $\omega^2(\tau)=0$ and reducing the Ermakov equation to
\begin{equation}\label{Ermakov}
\ddot{\rho} = \frac{b_0^2}{\rho^3},
\end{equation}
where $b_0$ is fixed by the initial conditions. Choosing $\rho(0)=1$ and $\dot{\rho}(0)=0$, the solution is
\begin{equation}\label{Sol_rho}
\rho(\tau) = \sqrt{1 + b_0^2 \tau^2}.
\end{equation}
This choice simplifies the analysis, although different initial conditions may be imposed depending on the physical configuration, such as applying the initial transformation $\hat{S}^\dagger(0)\hat{G}^\dagger(0)E(\mathbf{r}_\perp,0)$. Substituting the identity $\ddot{\rho}=b_0^2/\rho^3$ into Eq.~\eqref{H_S} gives the effective evolution equation
\begin{equation}\label{varphi}
\begin{split}
    i\frac{\partial}{\partial \tau}E_{S}(\mathbf{r}_\perp, \tau) &=\hat{H}_S(\tau)\,E_S(\mathbf{r}_\perp, \tau),\\
    &=\frac{1}{2\rho^2}\left(\hat{p}_x^2 + b_0^2 x^2 + \hat{p}_y^2 + b_0^2 y^2\right) E_S(\mathbf{r}_\perp, \tau),\\
    &=\frac{1}{2\rho^2}\sum_{\mu=x,y}\left(\hat{p}_\mu^2 + b_0^2 \mu^2\right)E_S(\mathbf{r}_\perp, \tau), 
\end{split}
\end{equation}
which corresponds to two decoupled harmonic oscillators with time-dependent scaling. This formulation is central in identifying mode structures and their invariants. In particular, the Hamiltonian in Eq.~\eqref{varphi} admits separation of variables in rectangular coordinates, making it directly suitable for describing Hermite-Gauss modes, which are standard solutions of the harmonic oscillator in optical systems. Although the Hamiltonian depends explicitly on the propagation parameter through $\rho(\tau)$, it commutes with itself at different values of $\tau$ \cite{Sakurai_2017,Ramos_2023a}. More importantly, Eq.~\eqref{varphi} establishes a direct equivalence between the paraxial wave equation and the propagation of optical beams in a quadratic GRIN-like medium. Thus, any known solution in a GRIN medium can be mapped onto a solution of Eq.~\eqref{Eq_1} via the inverse application of $\hat{G}(\tau)$ and $\hat{S}(\tau)$. This correspondence provides both a conceptual and a practical bridge between structured light in free space and its dynamics in GRIN optical media, forming a key result of this work.

Based on Eq.~\eqref{varphi}, the optical field in the transformed frame evolves as
\begin{equation}
    E_S(\mathbf{r}_\perp, \tau) = \hat{\mathcal{U}}(\tau)E_S(\mathbf{r}_\perp, 0),
\end{equation}
where $E_S(\mathbf{r}_\perp, 0)$ is the initial field in this frame and $\hat{\mathcal{U}}(\tau)$ is the evolution operator
\begin{equation}
    \begin{split}
        \hat{\mathcal{U}}(\tau) & = \exp\left(-i\frac{\alpha(\tau)}{2}\left(\hat{p}_x^2 + b_0^2 x^2 + \hat{p}_y^2 + b_0^2 y^2\right)\right), \\
        \alpha(\tau) & = \int_0^{\tau} \frac{dt}{\rho^2(t)} = \frac{1}{b_0}\arctan(b_0\tau),
    \end{split}
\end{equation}
with $\alpha(\tau)$ later identified as the accumulated Gouy phase~\cite{Berry_1984,Simon_1993}. The dynamics corresponds to two uncoupled harmonic oscillators, with all propagation effects contained in $\rho(\tau)$. Returning to the original frame requires the inverse transformations, yielding
\begin{equation}\label{ket_E}
    E(\mathbf{r}_\perp, \tau) = \hat{G}(\tau)\hat{S}(\tau)\hat{\mathcal{U}}(\tau)E(\mathbf{r}_\perp, 0),
\end{equation}
where $\hat{G}^\dagger(0)$ and $\hat{S}^\dagger(0)$ are omitted because the initial conditions $\rho(0)=1$ and $\dot{\rho}(0)=0$ ensure $E_S(\mathbf{r}_\perp, 0) = E(\mathbf{r}_\perp, 0)$. Eq.~\eqref{ket_E} expresses paraxial beam propagation through an effective oscillator model determined by $\rho(\tau)$: the operator $\hat{\mathcal{U}}(\tau)$ describes the evolution on a fixed harmonic oscillator basis, $\hat{S}(\tau)$ introduces the propagation-dependent scaling $\rho(\tau)=\sqrt{1+b_0^2\tau^2}$, and $\hat{G}(\tau)$ accounts for the quadratic phase associated with wavefront curvature. This operator representation provides a unified description of beam evolution, allowing the construction of Hermite-Gauss, Laguerre-Gauss, and Ince-Gauss modes, among others, through the spectral decomposition of $\hat{\mathcal{U}}(\tau)$ in the harmonic oscillator basis.
\subsection{Rectangular coordinates}
Once the general expression for the evolution of the optical field has been established in Eq.~\eqref{ket_E}, a natural starting point for the construction of explicit solutions is the rectangular coordinate system ($x,y,\tau=z/k$). In this case, the effective Hamiltonian associated with the evolution operator $\hat{\mathcal{U}}(\tau)$ takes the form (see Eq.~\eqref{varphi})
\begin{equation}\label{H_RC}
\hat{H}_{\texttt{RC}} = -\frac{1}{2}\left(\frac{\partial^2}{\partial x^2}+\frac{\partial^2}{\partial y^2}-b_0^2 x^2  - b_0^2 y^2\right).
\end{equation}
This corresponds to two uncoupled quantum harmonic oscillators along the $x$ and $y$ directions, both with the same frequency $b_0$. The eigenfunctions of this Hamiltonian operator, expressed in the number basis, are separable and take the form $\Psi_{n,m}(x,y) = \Psi_n(x)\Psi_m(y)$, where $n, m \in \mathbb{N}$. Each function $\Psi_n(\mu)$ satisfies the following
\begin{equation}\label{E_Hnm}
    \Psi_n(\mu) = \frac{(b_0/\pi)^{1/4}}{\sqrt{2^{n}n!}} \exp\left(-\frac{b_0 \mu^2}{2}\right) H_n\left(\sqrt{b_0} \mu\right),
\end{equation}
with $\mu = x, y$, and $H_n(\mu)$ denotes the $n$-th Hermite polynomial \cite{Sakurai_2017,Laser_Beam_Propagation}. The eigenvalue equation $\hat{H}_{\texttt{RC}} \Psi_{n,m}(x,y) = b_0(n + m + 1)\Psi_{n,m}(x,y)$ indicates that the functions $\Psi_{n,m}(x,y)$ constitute a complete orthonormal basis of solutions to the quadratic Hamiltonian defined in Eq.~\eqref{H_RC}. More importantly, this property allows for a straightforward application of the evolution operator $\hat{\mathcal{U}}(\tau)$. If the initial condition in Eq.~\eqref{ket_E} is taken to be one of these modes, that is, $E(\mathbf{r}_\perp, 0) = \Psi_{n,m}(x,y)$ (or a superposition thereof), then the solution at any propagation distance $\tau$ reads
\begin{equation}
\begin{split}
    E(\mathbf{r}_\perp, \tau) = &  \hat{G}(\tau)\hat{S}(\tau)\exp\left(-i\alpha(\tau) \hat{H}_{\texttt{RC}}\right) \Psi_{n,m}(x,y) \\
    =& \exp\left(-i\alpha(\tau)b_0(n + m + 1)\right) \\&\times\hat{G}(\tau)\hat{S}(\tau)\Psi_{n,m}(x,y),
\end{split}
\end{equation}
where the exponential factor accounts for the phase accumulated during propagation. The scaling transformation $\hat{S}(\tau)$ acts on the mode functions by rescaling their spatial arguments as: $\hat{S}(\tau)\Psi_{n,m}(x,y)\hat{S}^\dagger(\tau) = \Psi_{n,m}\left(x/\rho,y/\rho\right)$, which reflects the transverse expansion or compression of the beam profile. On the other hand, the transformation $\hat{G}(\tau)$ introduces a quadratic phase factor that depends on the propagation distance and accounts for the curvature of the wavefront. Together, these transformations describe the complete propagation dynamics of structured optical fields in free space. Therefore, the optical field $E(\mathbf{r}_\perp,\tau)$ evolves as follows
\begin{equation}\label{HGB}
    \begin{split}
        E(\mathbf{r}_\perp,\tau)& = A_{n,m}\frac{\exp\left(-i\alpha(\tau)b_0(n+m+1)\right)}{\rho}\\
        & \times \exp\left(i\frac{\dot{\rho}}{2\rho}\left(x^2+y^2\right)\right)\exp\left(-\frac{b_0}{2\rho^2}\left(x^2+y^2\right)\right) \\
        &\times H_n\left(\sqrt{b_0}\frac{x}{\rho}\right)H_m\left(\sqrt{b_0}\frac{y}{\rho}\right),
    \end{split}
\end{equation}
where $A_{n,m}$ is a normalization constant. 

\begin{figure}[h!]
\centering
\includegraphics[width=\linewidth]{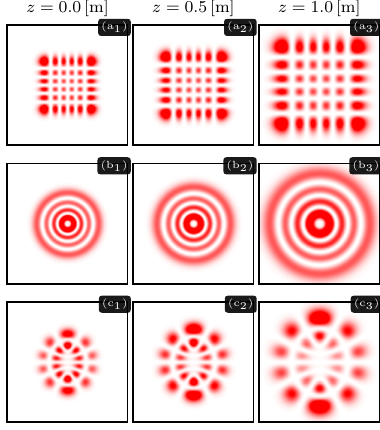}
  \caption{Transverse intensity distributions of paraxial beams at different propagation distances $z$. Subfigures (a$_1$–a$_3$) correspond to Hermite–Gauss beams with indices $(n,m)=(5,5)$ at $z = 0.0$ m, $0.5$ m, and $1.0$ m, respectively. Subfigures (b$_1$–b$_3$) show Laguerre–Gauss beams with $(n,m)=(3,2)$ at the same propagation distances, while (c$_1$–c$_3$) display the corresponding patterns for Ince–Gauss beams with $(n,m)=(7,5)$. Each case highlights the characteristic transverse profiles of the beam families, which are preserved during propagation up to a scaling transformation. The simulations are performed using red light with wavelength $\lambda = 633$ nm (He–Ne laser), an initial waist $W_0 \approx 200~\mu$m ($b_0=2/W_0^2$), and a Rayleigh range $z_R \approx 0.20$ m. Each case highlights the characteristic transverse profiles of the beam families, which are preserved during propagation up to a scaling transformation.}
\label{fig_1}
\end{figure}

To contextualize our results, we express the parameter $b_0$ in Eq.~\eqref{HGB} in terms of the standard Hermite–Gauss beam parameters, identifying $b_0 = 2/W_0^2$, where $W_0$ is the beam waist and $z_R = \pi W_0^2/\lambda$ is the Rayleigh range. Since $\tau = z/k$, and using Eq.~\eqref{Sol_rho} together with the previous relations, we get
\begin{equation}\label{relaciones}
\begin{split}
b_0 \alpha(\tau) &= \arctan(z/z_R),\\
\frac{\dot{\rho}}{\rho} &= \frac{k z}{z^2+z_R^2},\\
\end{split}
\qquad
\begin{split}
\frac{b_0}{2\rho^2} &= \frac{1}{W_0^2 \left(1+z^2/z_R^2\right)},\\
\frac{\sqrt{b_0}}{\rho} &= \frac{\sqrt{2}}{1+z^2/z_R^2}.
\end{split}
\end{equation}
As illustrated in Fig.~\ref{fig_1} (a$_1$, a$_2$, and a$_3$), which shows the evolution of Hermite–Gauss beams for selected parameter values and different propagation distances $z$, this formulation establishes a consistent framework for structured beam generation through distance-dependent unitary transformations derived from quantum optics. The agreement with the classical results \cite{Saleh_1991,Laser_Beam_Propagation} confirms the validity of the method and provides an alternative perspective to analyze these optical fields in terms of unitary transformations and operator algebra.

\subsection{Circular-cylindrical coordinates}
As a second case, we consider the circular-cylindrical coordinate system $(r, \theta, \tau = z/k)$. In this representation, the Hamiltonian associated with Eq.~\eqref{varphi}, formulated in transverse coordinates $(r, \theta)$, is given by the following:
\begin{equation}\label{H_CC}
    \hat{H}_{\texttt{CC}} = -\frac{1}{2}\left(\frac{\partial^2}{\partial r^2} + \frac{1}{r}\frac{\partial}{\partial r} +\frac{1}{r^2}\frac{\partial^2}{\partial\theta^2} - b_0^2 r^2\right),
\end{equation}
where $b_0$ is a constant parameter. This form of the Hamiltonian is particularly useful in systems with radial symmetry, such as optical fields propagating in waveguides or laser beams in circular-cylindrical coordinates \cite{He_1995,Oneil_2002,Plick_2015}. The eigenvalue equation for $\hat{H}_{\texttt{CC}}$ is given by: $\hat{H}_{\texttt{CC}} R_n(r)\Theta_m(\theta) = b_0(2n + |m| +1)R_n(r)\Theta_m(\theta)$, where $R_{n}(r)$ is the radial part of the optical field, and $\Theta_m(\theta) = \exp{(im\theta})$ describes the angular dependence. The eigenfunctions take the explicit form:
\begin{equation}
\begin{split}
    R_{n}(r)\Theta_m(\theta) &= A_{n,m}r^{|m|}\exp\left(-\frac{b_0r^2}{2}\right)\\
    &\times L_n^{|m|}\left(b_0r^2\right)\exp(im\theta),
    \end{split}
\end{equation}
where $A_{n,m}$ is a normalization constant, and $L_n^{|m|}\left(b_0r^2\right)$ are the associated Laguerre polynomials. This form captures the behavior of the system in polar coordinates, with the radial function depending on both the quantum numbers $n\in\mathbb{N}$ and $l\in\mathbb{Z}$, as well as the parameter $b_0$ \cite{Oneil_2002}. The evolution of the optical field, starting from the initial condition $E(\mathbf{r}_\perp,0)=R_n(r)\Theta_m(\theta)$, can be described through the action of the evolution operator defined in Eq.~\eqref{ket_E}, together with the Hamiltonian $\hat{H}_{\texttt{CC}}$ from Eq.~\eqref{H_CC}, and its corresponding eigenvalue equation. The optical field at a distance $\tau$ takes the form
\begin{equation}
\begin{split}
    E(\mathbf{r}_\perp,\tau) = & \hat{G}(\tau)\hat{S}(\tau)\exp\left(-i\alpha(\tau)\hat{H}_{\texttt{CC}}\right) R_n(r)\Theta_m(\theta)\\
    = & \exp\left(-i\alpha(\tau)b_0(2n+|m|+1)\right)\\
    &\times \hat{G}(\tau)\hat{S}(\tau) R_n(r)\Theta_m(\theta).
\end{split}
\end{equation}
Using the scaling identity $\hat{S}^\dagger(\tau) R_n(r) \hat{S}(\tau) = R_n(r/\rho)$, the evolved optical field simplifies to
\begin{equation}\label{LaguerreGauss}
\begin{split}
    E(\mathbf{r}_\perp,\tau) = & A_{n,m}\left(\frac{r}{\rho}\right)^{|m|}\exp\left(-i\alpha(\tau)b_0(2n+|m|+1)\right) \\
    &\times \exp\left(i\frac{\dot{\rho}}{2\rho}r^2\right) \exp\left(-\frac{b_0 r^2}{2\rho^2}\right) \\
    &\times L_n^{|m|}\left(\frac{b_0 r^2}{\rho^2}\right) \exp(im\theta),
\end{split}
\end{equation}
where $A_{n,m}$ is a normalization constant.

Eq.~\eqref{LaguerreGauss} describes the evolution of Laguerre-Gauss beams in circular–cylindrical coordinates, where the transverse scaling factor $\rho(\tau)$ determines the beam expansion, the phase term $\exp\left(i\frac{\dot{\rho}}{2\rho}r^2\right)$ accounts for the wavefront curvature, the radial profile is governed by the Laguerre polynomial $L_n^{|m|}(b_0 r^2/\rho^2)$, and the azimuthal dependence is given by $\exp(im\theta)$, in full agreement with the standard formulation of Laguerre-Gauss beams and following, analogously to the previous case, the relations in Eq.\eqref{relaciones}, while Fig.~\ref{fig_1} (b$_1$, b$_2$, and b$_3$) displays the transverse intensity distributions at propagation distance $z$, exhibiting the characteristic doughnut-shaped profile preserved during propagation.
\subsection{Elliptic-cylindrical coordinates}
Finally, the third representative coordinate system considered is the cylindrical elliptic coordinate system $(\xi,\eta,\tau = z/k)$, defined by the transformations $x = f \cosh(\xi)\cos(\eta)$ and $y = f \sinh(\xi)\sin(\eta)$, where $f$ is a scaling parameter associated with the focal distance of the underlying confocal ellipses. In this coordinate system, the Hamiltonian corresponding to Eq.~\eqref{ket_E} is expressed as \cite{Krenn_2013,Plick_2013}:
\begin{equation} 
    \begin{split}
    \hat{H}_{\texttt{EC}} =& -\frac{1}{2}\bigg[\frac{2}{f^2(\cosh 2\xi -\cos 2\eta)}\bigg(\frac{\partial^2}{\partial \xi^2} + \frac{\partial^2}{\partial \eta^2}\bigg)\\
    & -  b_0^2 f^2 (\cosh 2\xi + \cos 2\eta)\bigg], 
    \end{split}
\end{equation}
and the associated eigenvalue equation under the separation of variables is $\hat{H}_{\texttt{EC}} U(\xi)V(\eta) = b_0(p+1)U(\xi)V(\eta)$, whose solutions can be written as
\begin{equation}
\begin{split}
  U(\xi)V(\eta) = & \mathcal{C}_p^m(i\xi,\epsilon)\mathcal{C}_p^m(\eta,\epsilon)\\
  &\times\exp\left(-\frac{\epsilon}{4}(\cosh 2\xi + \cos 2\eta)\right),
  \end{split}
\end{equation}
with $\epsilon = b_0f^2$ the ellipticity parameter. These eigenfunctions are expressed in terms of the even and odd Ince polynomials $\mathcal{C}_p^m$ and $\mathcal{S}_p^m$, respectively, where $p$ denotes the order and $m$ the degree of the polynomial, with $0 \leq m \leq p$ for even solutions and $1 \leq m \leq p$ for odd solutions \cite{Boyer_1975}. Hence, the optical field in this coordinate system evolves as
\begin{equation}
     \begin{split}
         E(\mathbf{r}_\perp,\tau) = & \hat{G}(\tau)\hat{S}(\tau)\exp\left(-i\alpha(\tau)\hat{H}_{\texttt{EC}}\right) U(\xi)V(\eta)\\
         =&\exp\left(-i\alpha(\tau)b_0(p+1)\right)
         \\&\times\hat{G}(\tau)\hat{S}(\tau)\mathcal{C}_p^m\left(i\xi,\epsilon\right)\mathcal{C}_p^m\left(\eta,\epsilon\right)\\
         &\times\exp\left(-\frac{\epsilon}{4}(\cosh 2\xi + \cos 2\eta)\right),
     \end{split}
\end{equation}
and, after applying the similarity transformations, the field propagates as
\begin{equation}\label{InceGauss}
    \begin{split}
          E(\mathbf{r}_\perp, \tau) = & \frac{\exp\left(i\frac{\dot{\rho}}{4\rho}f^2(\cosh{2\xi} + \cos{2\eta})\right)}{\rho}\\
          &\times\exp[-i\alpha(\tau)b_0(p+1)]\mathcal{C}_p^m(\eta',\epsilon)\mathcal{C}_p^m(i\xi', \epsilon)\\
          &\times \exp\left(-\frac{\epsilon}{4} \frac{\cosh{2\xi} + \cos{2\eta}}{\rho^2}\right),
    \end{split}
\end{equation}
with analogous expressions for the odd eigenfunctions $\mathcal{S}_{p}^{m}(\eta,\epsilon)\mathcal{S}_{p}^{m}(i\xi,\epsilon)$. Here, we have used the transformation property $\hat{S}^{\dagger}(\tau)f(\xi,\eta)\hat{S}(\tau) = f(\xi',\eta')$, where $\xi' = \xi(x/\rho, y/\rho)$ and $\eta' = \eta(x/\rho, y/\rho)$.

Finally, and analogously to the two representative cases discussed above, we further contextualize our results through unitary transformations by analyzing Eq.~\eqref{InceGauss}, where the scaling factor $\rho(\tau)$ governs the transverse expansion of the beam, and the phase term $\exp\left[i\frac{\dot{\rho}}{4\rho}f^2(\cosh{2\xi} + \cos{2\eta})\right]$ accounts for the wavefront curvature during propagation. The transverse intensity distribution is determined by the Ince polynomials $\mathcal{C}_p^m(i\xi', \epsilon)$ and $\mathcal{C}_p^m(\eta', \epsilon)$, while the azimuthal dependence is described by the phase factor $\exp(im\eta)$. This formulation is consistent with the standard description of Ince–Gauss beams and, taking into account the relations in Eq.~\eqref{relaciones} together with the experimental values of the parameters, Fig.~\ref{fig_1} (c$_1$, c$_2$, and c$_3$) illustrates the transverse intensity pattern at different propagation distances $z$, showing the characteristic elliptical profile preserved during propagation. 
\section{Lewis-Ermakov invariant}\label{LEI}
Finally, we develop the Lewis-Ermakov invariant in the context of the paraxial wave equation and discuss its physical implications. We analyze the instantaneous eigenvalues and eigenfunctions of the effective Hamiltonian, which are essential for understanding the dynamics and evolution of structured optical beams during propagation. In particular, we examine how the presence of a Gaussian envelope causes any finite-energy beam to propagate as if it were subject to a quadratic GRIN-like potential. The unitary transformation approach clarifies how the Gaussian envelope of physical beams induces effective harmonic confinement and connects the propagation dynamics to oscillator-like invariants.

To establish this connection, we investigate the relationship between the paraxial wave equation, Eq.~\eqref{Eq_1}, the Lewis-Ermakov invariant, and the evolution of optical fields in a quadratic GRIN-like medium, as described by the unitary operators $\hat{G}(\tau)$ and $\hat{S}(\tau)$. By applying these transformations successively to the Hamiltonian in Eq.~\eqref{varphi} and evaluating at $\tau=0$, we obtain the following
\begin{equation}\label{H_S0}
    \hat{H}_S(0)=\frac{1}{2\rho(0)^2}\sum_{\mu=x,y}\left(\hat{p}_\mu^2+b_0^2\mu^2\right).
\end{equation}
Setting $\rho(0)=1$ for simplicity, the Lewis-Ermakov invariant can be written as \cite{Leach_1977a,Leach_1977b}:
\begin{equation}
\begin{split}\label{Invariante}
    \hat{I}(\tau) &= \hat{G}(\tau)\hat{S}(\tau)\hat{H}_S(0)\hat{S}^\dagger(\tau)\hat{G}^\dagger(\tau),\\
    &= \frac{1}{2}\sum_{\mu=x,y}\left[\left(\rho\hat{p}_\mu-\dot{\rho}\mu\right)^2+\frac{b_0^2}{\rho^2}\mu^2\right].
\end{split}
\end{equation}
For clarity, we perform the analysis in rectangular coordinates; however, the same procedure can be applied in any other coordinate system that incorporates a Gaussian envelope. This choice allows for an explicit verification of the standard condition for a dynamical invariant:
\begin{equation}\label{Eq_I}
\frac{d\hat{I}}{d\tau} = \frac{\partial \hat{I}}{\partial \tau} - i[\hat{I},\hat{H}_{\texttt{free}}] = 0,
\end{equation}
where $\hat{H}_{\texttt{free}} = (\hat{p}_x^2+\hat{p}_y^2)/2$ is the Hamiltonian in free space introduced in Eq.~\eqref{Eq_1}. Substituting the auxiliary condition $\ddot{\rho}=b_0^2/\rho^3$, it follows that Eq.~\eqref{Eq_I} is identically satisfied. Thus, the Lewis-Ermakov invariant defines an operator that remains conserved during propagation, expressed in terms of canonical variables and their derivatives.

A key aspect of this construction is that, at $\tau=0$, the invariant $\hat{I}(0)$ and the free-space Hamiltonian $\hat{H}_{\texttt{free}}$ do not commute and therefore do not share a common eigenbasis. Specifically, from Eq.~\eqref{Invariante} with $\rho(0)=1$ and $\dot{\rho}(0)=0$, one obtains
\begin{equation}
    \hat{I}(0) = \frac{1}{2} \sum_{\mu = x, y} \left(\hat{p}_\mu^2 + b_0^2 \mu^2\right),
\end{equation}
which corresponds to the Hamiltonian of a two-dimensional harmonic oscillator with frequency $b_0$, in contrast to the Hamiltonian of a free particle $\hat{H}_{\texttt{free}}=(\hat{p}_x^2+\hat{p}_y^2)/2$. This difference arises from the Gaussian modulation $\exp[-b_0(x^2+y^2)/2]$, which introduces a quadratic confinement. Since $b_0\propto 1/W_0^2$, exact commutativity at $\tau=0$ would require $b_0=0$, corresponding to an infinitely wide nonsquare integrable beam. As a result, any finite-energy beam propagates as if it were subject to a quadratic GRIN-like potential. The unitary transformation approach makes this distinction explicit, showing how the Gaussian envelope of physical beams induces a harmonic confinement that connects free-space propagation with oscillator invariants. This result, together with the operator-based approach and the use of unitary transformations in optical fields governed by the paraxial wave equation, Eq.~\eqref{Eq_1}, constitutes a central contribution of this work.
\subsection{Instantaneous eigenvalues and eigenfunctions}
We now turn our attention to the eigenvalues and eigenfunctions of the Hamiltonian $\hat{H}_S$. Through the application of the transformations $\hat{G}(\tau)$ and $\hat{S}(\tau)$, it can be demonstrated that although there is a formal connection to the Hamiltonian in the free space $\hat{H}_{\texttt{free}}$, the resulting eigenvalues and eigenfunctions do not coincide with those of $\hat{H}_{\texttt{free}}$. This discrepancy arises from the presence of dynamical terms that are intrinsic to the propagation of states in time-dependent systems and must be properly accounted for. In the context of the paraxial wave equation, this situation is well understood within the framework of shortcuts to adiabaticity \cite{Berry_2009,Martinez_2017,Guery_2019,Kong_2020,Li_2024}.

To ensure clarity, we performed our analysis using rectangular coordinates. We start by examining the eigenfunctions $\Psi_n(x)\Psi_m(y)$, as defined in Eq.~\eqref{E_Hnm}, which, in the final transformed frame (see Eq.~\eqref{varphi}), satisfy
\begin{equation}\label{eigen_H_S}
    \hat{H}_{S}(\tau)\,\Psi_{n}(x)\Psi_m(y)= \frac{b_0(n + m + 1)}{\rho^2}\Psi_{n}(x)\Psi_m(y).
\end{equation}
To relate instantaneous eigenvalues and eigenfunctions in the original frame, recall that $\hat{H}_S = \hat{S}^\dagger\hat{H}_G\hat{S}-i\hat{S}^\dagger\frac{\partial\hat{S}}{\partial\tau}$. Multiplying Eq.~\eqref{eigen_H_S} from the left by $\hat{S}(\tau)$ yields
\begin{equation}
\begin{split}
    &\left[\hat{H}_G(\tau)-\frac{\dot{\rho}}{2\rho}\sum_{\mu=x,y}\left(\mu\hat{p}_\mu+\hat{p}_\mu\mu\right)\right]\hat{S}(\tau)\Psi_{n}(x)\Psi_{m}(y)\\
    &=\frac{b_0(n + m + 1)}{\rho^2}\hat{S}(\tau)\Psi_{n}(x)\Psi_m(y).
\end{split}
\end{equation}
Furthermore, since $\hat{H}_G=\hat{G}^\dagger\hat{H}_\texttt{free}\hat{G}-i\hat{G}^\dagger\frac{\partial\hat{G}}{\partial\tau}$, multiplying from the left by $\hat{G}(\tau)$ gives
\begin{equation}
    \begin{split}
    &\bigg\{\hat{H}_\texttt{free}+\frac{1}{2}\sum_{\mu=x,y}\bigg[\left(\frac{\ddot{\rho}\rho+\dot{\rho}^2}{\rho^2}\right)\mu^2\\&-\frac{\dot{\rho}}{\rho}\left(\mu\hat{p}_\mu+\hat{p}_\mu\mu\right)\bigg]\bigg\}\hat{G}(\tau)\hat{S}(\tau)\Psi_{n}(x)\Psi_m(y)\\
    &=\frac{b_0(n + m + 1)}{\rho^2}\hat{G}(\tau)\hat{S}(\tau)\Psi_{n}(x)\Psi_m(y).
    \end{split}
\end{equation}
This eigenvalue equation shows that the eigenvalues are instantaneous, since they explicitly depend on $1/\rho^2$, and the corresponding instantaneous eigenfunctions are given by $\hat{G}(\tau)\hat{S}(\tau)\Psi_n(x)\Psi_m(y)$. Importantly, these eigenvalues do not correspond to those of $\hat{H}_\texttt{free}$, but rather to those of a Hamiltonian with two additional terms responsible for generating the dynamical phase and implementing the rescaling of the beam profiles. Moreover, this framework naturally accommodates the inclusion of adiabatic-invariant protocols through the Hamiltonian structure, enabling the systematic addition of potentials and compensating terms designed to suppress propagation-induced effects such as mode coupling or excitation of unwanted modes.

\section{Conclusions}\label{Conclusiones}
We develop an operator-based framework for analyzing the propagation of structured optical beams in free space, formulated through a sequence of propagation-dependent unitary transformations. By recasting the paraxial wave equation in this language, the beam dynamics is mapped onto those of two uncoupled harmonic oscillators with propagation-dependent parameters. The introduction of the auxiliary scaling function $\rho(\tau)$, governed by the Ermakov equation, provides a compact and unified description of the field evolution, simultaneously accounting for transverse scaling and accumulated phase effects. Within this formalism, families of modes such as Hermite-Gauss, Laguerre-Gauss, and Ince-Gauss naturally emerge as eigenfunctions of the effective Hamiltonian, particularly in coordinate systems that allow the separation of variables. This structure offers a systematic route to constructing exact solutions and interpreting the behavior of structured beams, while clarifying the role of propagation invariants, underlying symmetries, and the physical significance of the Gaussian envelope as an effective quadratic confinement.

Furthermore, the framework establishes a direct connection between the Lewis-Ermakov invariant and the propagation of finite-energy beams, demonstrating that any physically realizable beam evolves as if subject to a quadratic GRIN-like potential. The non-commutativity between the dynamical invariant and the free-space Hamiltonian at the initial stage, arising from the Gaussian envelope, underscores the fundamental distinction between idealized and physically realizable optical fields.

Beyond its optical relevance, the formalism applies directly to quantum systems governed by time-dependent quadratic Hamiltonians. The evolution of a free particle or a particle in a two-dimensional harmonic potential can be described using the same unitary transformations and auxiliary function $\rho(\tau)$, recovering established results associated with Lewis-Riesenfeld invariants while offering new insights into wave packet dynamics. The structural equivalence between optical beam propagation and quantum evolution highlights the broad applicability of this operator-based approach, providing a unified language for exploring both structured light and fundamental problems in quantum mechanics involving time-dependent or spatially modulated potentials.


\begin{thebibliography}{48}%
\makeatletter
\providecommand \@ifxundefined [1]{%
 \@ifx{#1\undefined}
}%
\providecommand \@ifnum [1]{%
 \ifnum #1\expandafter \@firstoftwo
 \else \expandafter \@secondoftwo
 \fi
}%
\providecommand \@ifx [1]{%
 \ifx #1\expandafter \@firstoftwo
 \else \expandafter \@secondoftwo
 \fi
}%
\providecommand \natexlab [1]{#1}%
\providecommand \enquote  [1]{``#1''}%
\providecommand \bibnamefont  [1]{#1}%
\providecommand \bibfnamefont [1]{#1}%
\providecommand \citenamefont [1]{#1}%
\providecommand \href@noop [0]{\@secondoftwo}%
\providecommand \href [0]{\begingroup \@sanitize@url \@href}%
\providecommand \@href[1]{\@@startlink{#1}\@@href}%
\providecommand \@@href[1]{\endgroup#1\@@endlink}%
\providecommand \@sanitize@url [0]{\catcode `\\12\catcode `\$12\catcode `\&12\catcode `\#12\catcode `\^12\catcode `\_12\catcode `\%12\relax}%
\providecommand \@@startlink[1]{}%
\providecommand \@@endlink[0]{}%
\providecommand \url  [0]{\begingroup\@sanitize@url \@url }%
\providecommand \@url [1]{\endgroup\@href {#1}{\urlprefix }}%
\providecommand \urlprefix  [0]{URL }%
\providecommand \Eprint [0]{\href }%
\providecommand \doibase [0]{https://doi.org/}%
\providecommand \selectlanguage [0]{\@gobble}%
\providecommand \bibinfo  [0]{\@secondoftwo}%
\providecommand \bibfield  [0]{\@secondoftwo}%
\providecommand \translation [1]{[#1]}%
\providecommand \BibitemOpen [0]{}%
\providecommand \bibitemStop [0]{}%
\providecommand \bibitemNoStop [0]{.\EOS\space}%
\providecommand \EOS [0]{\spacefactor3000\relax}%
\providecommand \BibitemShut  [1]{\csname bibitem#1\endcsname}%
\let\auto@bib@innerbib\@empty
\bibitem [{\citenamefont {Saleh}\ and\ \citenamefont {Teich}(1991)}]{Saleh_1991}%
  \BibitemOpen
  \bibfield  {author} {\bibinfo {author} {\bibfnamefont {B.~E.~A.}\ \bibnamefont {Saleh}}\ and\ \bibinfo {author} {\bibfnamefont {M.~C.}\ \bibnamefont {Teich}},\ }\href {https://doi.org/10.1002/0471213748} {\emph {\bibinfo {title} {Fundamentals of Photonics}}}\ (\bibinfo  {publisher} {Wiley},\ \bibinfo {year} {1991})\BibitemShut {NoStop}%
\bibitem [{\citenamefont {Forbes}(2014)}]{Laser_Beam_Propagation}%
  \BibitemOpen
  \bibinfo {editor} {\bibfnamefont {A.}~\bibnamefont {Forbes}},\ ed.,\ \href {https://doi.org/10.1201/b16548} {\emph {\bibinfo {title} {Laser Beam Propagation: Generation and Propagation of Customized Light}}},\ \bibinfo {edition} {1st}\ ed.\ (\bibinfo  {publisher} {CRC Press},\ \bibinfo {year} {2014})\BibitemShut {NoStop}%
\bibitem [{\citenamefont {Boyd}\ and\ \citenamefont {Gordon}(1961)}]{Boyd_HGB_1961}%
  \BibitemOpen
  \bibfield  {author} {\bibinfo {author} {\bibfnamefont {G.~D.}\ \bibnamefont {Boyd}}\ and\ \bibinfo {author} {\bibfnamefont {J.~P.}\ \bibnamefont {Gordon}},\ }\bibfield  {title} {\bibinfo {title} {Confocal multimode resonator for millimeter through optical wavelength masers},\ }\href {https://doi.org/https://doi.org/10.1002/j.1538-7305.1961.tb01626.x} {\bibfield  {journal} {\bibinfo  {journal} {Bell System Technical Journal}\ }\textbf {\bibinfo {volume} {40}},\ \bibinfo {pages} {489} (\bibinfo {year} {1961})}\BibitemShut {NoStop}%
\bibitem [{\citenamefont {Allen}\ \emph {et~al.}(1992)\citenamefont {Allen}, \citenamefont {Beijersbergen}, \citenamefont {Spreeuw},\ and\ \citenamefont {Woerdman}}]{Allen_LGB_1992}%
  \BibitemOpen
  \bibfield  {author} {\bibinfo {author} {\bibfnamefont {L.}~\bibnamefont {Allen}}, \bibinfo {author} {\bibfnamefont {M.~W.}\ \bibnamefont {Beijersbergen}}, \bibinfo {author} {\bibfnamefont {R.~J.~C.}\ \bibnamefont {Spreeuw}},\ and\ \bibinfo {author} {\bibfnamefont {J.~P.}\ \bibnamefont {Woerdman}},\ }\bibfield  {title} {\bibinfo {title} {Orbital angular momentum of light and the transformation of {L}aguerre-{G}aussian laser modes},\ }\href {https://doi.org/10.1103/PhysRevA.45.8185} {\bibfield  {journal} {\bibinfo  {journal} {Phys. Rev. A}\ }\textbf {\bibinfo {volume} {45}},\ \bibinfo {pages} {8185} (\bibinfo {year} {1992})}\BibitemShut {NoStop}%
\bibitem [{\citenamefont {Plick}\ and\ \citenamefont {Krenn}(2015)}]{Plick_2015}%
  \BibitemOpen
  \bibfield  {author} {\bibinfo {author} {\bibfnamefont {W.~N.}\ \bibnamefont {Plick}}\ and\ \bibinfo {author} {\bibfnamefont {M.}~\bibnamefont {Krenn}},\ }\bibfield  {title} {\bibinfo {title} {Physical meaning of the radial index of {L}aguerre-{G}auss beams},\ }\href {https://doi.org/10.1103/PhysRevA.92.063841} {\bibfield  {journal} {\bibinfo  {journal} {Phys. Rev. A}\ }\textbf {\bibinfo {volume} {92}},\ \bibinfo {pages} {063841} (\bibinfo {year} {2015})}\BibitemShut {NoStop}%
\bibitem [{\citenamefont {Mardoyan}\ \emph {et~al.}(1985)\citenamefont {Mardoyan}, \citenamefont {Pogosyan}, \citenamefont {Sissaklan},\ and\ \citenamefont {Ter-Antonyan}}]{Mardoyan1985}%
  \BibitemOpen
  \bibfield  {author} {\bibinfo {author} {\bibfnamefont {L.~G.}\ \bibnamefont {Mardoyan}}, \bibinfo {author} {\bibfnamefont {G.~S.}\ \bibnamefont {Pogosyan}}, \bibinfo {author} {\bibfnamefont {A.~N.}\ \bibnamefont {Sissaklan}},\ and\ \bibinfo {author} {\bibfnamefont {V.~M.}\ \bibnamefont {Ter-Antonyan}},\ }\bibfield  {title} {\bibinfo {title} {Elliptic basis of circular oscillator},\ }\href {https://doi.org/10.1007/BF02729028} {\bibfield  {journal} {\bibinfo  {journal} {Il Nuovo Cimento B (1971-1996)}\ }\textbf {\bibinfo {volume} {88}},\ \bibinfo {pages} {43} (\bibinfo {year} {1985})}\BibitemShut {NoStop}%
\bibitem [{\citenamefont {Bandres}\ and\ \citenamefont {Guti\'{e}rrez-Vega}(2004)}]{Bandres_Ince_2004}%
  \BibitemOpen
  \bibfield  {author} {\bibinfo {author} {\bibfnamefont {M.~A.}\ \bibnamefont {Bandres}}\ and\ \bibinfo {author} {\bibfnamefont {J.~C.}\ \bibnamefont {Guti\'{e}rrez-Vega}},\ }\bibfield  {title} {\bibinfo {title} {Ince-{G}aussian beams},\ }\href {https://doi.org/10.1364/OL.29.000144} {\bibfield  {journal} {\bibinfo  {journal} {Opt. Lett.}\ }\textbf {\bibinfo {volume} {29}},\ \bibinfo {pages} {144} (\bibinfo {year} {2004})}\BibitemShut {NoStop}%
\bibitem [{\citenamefont {Lewis}(1967)}]{Lewis_1967}%
  \BibitemOpen
  \bibfield  {author} {\bibinfo {author} {\bibfnamefont {H.~R.}\ \bibnamefont {Lewis}},\ }\bibfield  {title} {\bibinfo {title} {Classical and quantum systems with time-dependent harmonic-oscillator-type hamiltonians},\ }\href {https://doi.org/10.1103/PhysRevLett.18.510} {\bibfield  {journal} {\bibinfo  {journal} {Phys. Rev. Lett.}\ }\textbf {\bibinfo {volume} {18}},\ \bibinfo {pages} {510} (\bibinfo {year} {1967})}\BibitemShut {NoStop}%
\bibitem [{\citenamefont {Lewis}\ and\ \citenamefont {Riesenfeld}(1969)}]{Lewis_1969}%
  \BibitemOpen
  \bibfield  {author} {\bibinfo {author} {\bibfnamefont {J.}~\bibnamefont {Lewis}, \bibfnamefont {H.~R.}}\ and\ \bibinfo {author} {\bibfnamefont {W.~B.}\ \bibnamefont {Riesenfeld}},\ }\bibfield  {title} {\bibinfo {title} {{An Exact Quantum Theory of the Time‐Dependent Harmonic Oscillator and of a Charged Particle in a Time‐Dependent Electromagnetic Field}},\ }\href {https://doi.org/10.1063/1.1664991} {\bibfield  {journal} {\bibinfo  {journal} {Journal of Mathematical Physics}\ }\textbf {\bibinfo {volume} {10}},\ \bibinfo {pages} {1458} (\bibinfo {year} {1969})}\BibitemShut {NoStop}%
\bibitem [{\citenamefont {Leach}(1977{\natexlab{a}})}]{Leach_1977a}%
  \BibitemOpen
  \bibfield  {author} {\bibinfo {author} {\bibfnamefont {P.~G.~L.}\ \bibnamefont {Leach}},\ }\bibfield  {title} {\bibinfo {title} {{On the theory of time‐dependent linear canonical transformations as applied to Hamiltonians of the harmonic oscillator type}},\ }\href {https://doi.org/10.1063/1.523447} {\bibfield  {journal} {\bibinfo  {journal} {Journal of Mathematical Physics}\ }\textbf {\bibinfo {volume} {18}},\ \bibinfo {pages} {1608} (\bibinfo {year} {1977}{\natexlab{a}})}\BibitemShut {NoStop}%
\bibitem [{\citenamefont {Leach}(1977{\natexlab{b}})}]{Leach_1977b}%
  \BibitemOpen
  \bibfield  {author} {\bibinfo {author} {\bibfnamefont {P.~G.~L.}\ \bibnamefont {Leach}},\ }\bibfield  {title} {\bibinfo {title} {{Invariants and wavefunctions for some time‐dependent harmonic oscillator‐type Hamiltonians}},\ }\href {https://doi.org/10.1063/1.523161} {\bibfield  {journal} {\bibinfo  {journal} {Journal of Mathematical Physics}\ }\textbf {\bibinfo {volume} {18}},\ \bibinfo {pages} {1902} (\bibinfo {year} {1977}{\natexlab{b}})}\BibitemShut {NoStop}%
\bibitem [{\citenamefont {Markov}(1988)}]{Markov}%
  \BibitemOpen
  \bibfield  {author} {\bibinfo {author} {\bibfnamefont {M.~A.}\ \bibnamefont {Markov}},\ }\href {https://www.osti.gov/biblio/6509274} {\emph {\bibinfo {title} {Invariants and the evolution of nonstationary quantum system}}}\ (\bibinfo  {publisher} {Commack, NY (USA); Nova Science Publishers, Inc.},\ \bibinfo {year} {1988})\BibitemShut {NoStop}%
\bibitem [{\citenamefont {Dodonov}\ and\ \citenamefont {Man’ko}(1987)}]{Dodonov_1987}%
  \BibitemOpen
  \bibfield  {author} {\bibinfo {author} {\bibfnamefont {V.}~\bibnamefont {Dodonov}}\ and\ \bibinfo {author} {\bibfnamefont {V.}~\bibnamefont {Man’ko}},\ }\bibfield  {title} {\bibinfo {title} {Invariants and correlated states of nonstationary quantum systems},\ }\href@noop {} {\bibfield  {journal} {\bibinfo  {journal} {Invariants and the Evolution of Nonstationary Quantum Systems, Proceedings of Lebedev Physics Institute}\ }\textbf {\bibinfo {volume} {183}},\ \bibinfo {pages} {71} (\bibinfo {year} {1987})}\BibitemShut {NoStop}%
\bibitem [{\citenamefont {Dodonov}(2000)}]{Dodonov_2000}%
  \BibitemOpen
  \bibfield  {author} {\bibinfo {author} {\bibfnamefont {V.~V.}\ \bibnamefont {Dodonov}},\ }\bibfield  {title} {\bibinfo {title} {Universal integrals of motion and universal invariants of quantum systems},\ }\href {https://doi.org/10.1088/0305-4470/33/43/305} {\bibfield  {journal} {\bibinfo  {journal} {Journal of Physics A: Mathematical and General}\ }\textbf {\bibinfo {volume} {33}},\ \bibinfo {pages} {7721} (\bibinfo {year} {2000})}\BibitemShut {NoStop}%
\bibitem [{\citenamefont {Dodonov}\ and\ \citenamefont {Man'ko}(2000)}]{Dodonov_2000b}%
  \BibitemOpen
  \bibfield  {author} {\bibinfo {author} {\bibfnamefont {V.~V.}\ \bibnamefont {Dodonov}}\ and\ \bibinfo {author} {\bibfnamefont {O.~V.}\ \bibnamefont {Man'ko}},\ }\bibfield  {title} {\bibinfo {title} {Universal invariants of quantum-mechanical and optical systems},\ }\href {https://doi.org/10.1364/JOSAA.17.002403} {\bibfield  {journal} {\bibinfo  {journal} {J. Opt. Soc. Am. A}\ }\textbf {\bibinfo {volume} {17}},\ \bibinfo {pages} {2403} (\bibinfo {year} {2000})}\BibitemShut {NoStop}%
\bibitem [{\citenamefont {Ramos-Prieto}\ \emph {et~al.}(2018)\citenamefont {Ramos-Prieto}, \citenamefont {Urz{\'u}a-Pineda}, \citenamefont {Soto-Eguibar},\ and\ \citenamefont {Moya-Cessa}}]{Ramos_2018b}%
  \BibitemOpen
  \bibfield  {author} {\bibinfo {author} {\bibfnamefont {I.}~\bibnamefont {Ramos-Prieto}}, \bibinfo {author} {\bibfnamefont {A.~R.}\ \bibnamefont {Urz{\'u}a-Pineda}}, \bibinfo {author} {\bibfnamefont {F.}~\bibnamefont {Soto-Eguibar}},\ and\ \bibinfo {author} {\bibfnamefont {H.~M.}\ \bibnamefont {Moya-Cessa}},\ }\bibfield  {title} {\bibinfo {title} {{KvN} mechanics approach to the time-dependent frequency harmonic oscillator},\ }\href {https://doi.org/10.1038/s41598-018-26759-w} {\bibfield  {journal} {\bibinfo  {journal} {Scientific Reports}\ }\textbf {\bibinfo {volume} {8}},\ \bibinfo {pages} {8401} (\bibinfo {year} {2018})}\BibitemShut {NoStop}%
\bibitem [{\citenamefont {Urzúa}\ \emph {et~al.}(2019)\citenamefont {Urzúa}, \citenamefont {Ramos-Prieto}, \citenamefont {Fernández-Guasti},\ and\ \citenamefont {Moya-Cessa}}]{Urzua_2019}%
  \BibitemOpen
  \bibfield  {author} {\bibinfo {author} {\bibfnamefont {A.~R.}\ \bibnamefont {Urzúa}}, \bibinfo {author} {\bibfnamefont {I.}~\bibnamefont {Ramos-Prieto}}, \bibinfo {author} {\bibfnamefont {M.}~\bibnamefont {Fernández-Guasti}},\ and\ \bibinfo {author} {\bibfnamefont {H.~M.}\ \bibnamefont {Moya-Cessa}},\ }\bibfield  {title} {\bibinfo {title} {Solution to the time-dependent coupled harmonic oscillators hamiltonian with arbitrary interactions},\ }\href {https://doi.org/10.3390/quantum1010009} {\bibfield  {journal} {\bibinfo  {journal} {Quantum Reports}\ }\textbf {\bibinfo {volume} {1}},\ \bibinfo {pages} {82} (\bibinfo {year} {2019})}\BibitemShut {NoStop}%
\bibitem [{\citenamefont {Ramos-Prieto}\ \emph {et~al.}(2023)\citenamefont {Ramos-Prieto}, \citenamefont {Rom\'an-Ancheyta}, \citenamefont {Soto-Eguibar}, \citenamefont {R\'ecamier},\ and\ \citenamefont {Moya-Cessa}}]{Ramos_2023a}%
  \BibitemOpen
  \bibfield  {author} {\bibinfo {author} {\bibfnamefont {I.}~\bibnamefont {Ramos-Prieto}}, \bibinfo {author} {\bibfnamefont {R.}~\bibnamefont {Rom\'an-Ancheyta}}, \bibinfo {author} {\bibfnamefont {F.}~\bibnamefont {Soto-Eguibar}}, \bibinfo {author} {\bibfnamefont {J.}~\bibnamefont {R\'ecamier}},\ and\ \bibinfo {author} {\bibfnamefont {H.~M.}\ \bibnamefont {Moya-Cessa}},\ }\bibfield  {title} {\bibinfo {title} {Temporal factorization of a nonstationary electromagnetic cavity field},\ }\href {https://doi.org/10.1103/PhysRevA.108.033720} {\bibfield  {journal} {\bibinfo  {journal} {Phys. Rev. A}\ }\textbf {\bibinfo {volume} {108}},\ \bibinfo {pages} {033720} (\bibinfo {year} {2023})}\BibitemShut {NoStop}%
\bibitem [{\citenamefont {Nazarathy}\ and\ \citenamefont {Shamir}(1980)}]{Nazarathy:80}%
  \BibitemOpen
  \bibfield  {author} {\bibinfo {author} {\bibfnamefont {M.}~\bibnamefont {Nazarathy}}\ and\ \bibinfo {author} {\bibfnamefont {J.}~\bibnamefont {Shamir}},\ }\bibfield  {title} {\bibinfo {title} {Fourier optics described by operator algebra},\ }\href {https://doi.org/10.1364/JOSA.70.000150} {\bibfield  {journal} {\bibinfo  {journal} {J. Opt. Soc. Am.}\ }\textbf {\bibinfo {volume} {70}},\ \bibinfo {pages} {150} (\bibinfo {year} {1980})}\BibitemShut {NoStop}%
\bibitem [{\citenamefont {Stoler}(1981)}]{Stoler_1981}%
  \BibitemOpen
  \bibfield  {author} {\bibinfo {author} {\bibfnamefont {D.}~\bibnamefont {Stoler}},\ }\bibfield  {title} {\bibinfo {title} {Operator methods in physical optics},\ }\href {https://doi.org/10.1364/JOSA.71.000334} {\bibfield  {journal} {\bibinfo  {journal} {J. Opt. Soc. Am.}\ }\textbf {\bibinfo {volume} {71}},\ \bibinfo {pages} {334} (\bibinfo {year} {1981})}\BibitemShut {NoStop}%
\bibitem [{\citenamefont {Nienhuis}\ and\ \citenamefont {Allen}(1993)}]{Nienhuis_1993}%
  \BibitemOpen
  \bibfield  {author} {\bibinfo {author} {\bibfnamefont {G.}~\bibnamefont {Nienhuis}}\ and\ \bibinfo {author} {\bibfnamefont {L.}~\bibnamefont {Allen}},\ }\bibfield  {title} {\bibinfo {title} {Paraxial wave optics and harmonic oscillators},\ }\href {https://doi.org/10.1103/PhysRevA.48.656} {\bibfield  {journal} {\bibinfo  {journal} {Phys. Rev. A}\ }\textbf {\bibinfo {volume} {48}},\ \bibinfo {pages} {656} (\bibinfo {year} {1993})}\BibitemShut {NoStop}%
\bibitem [{\citenamefont {Moshinsky}(1973)}]{Moshinsky_1973}%
  \BibitemOpen
  \bibfield  {author} {\bibinfo {author} {\bibfnamefont {M.}~\bibnamefont {Moshinsky}},\ }\bibfield  {title} {\bibinfo {title} {Canonical transformations and quantum mechanics},\ }\href {https://doi.org/10.1137/0125024} {\bibfield  {journal} {\bibinfo  {journal} {SIAM Journal on Applied Mathematics}\ }\textbf {\bibinfo {volume} {25}},\ \bibinfo {pages} {193} (\bibinfo {year} {1973})},\ \Eprint {https://arxiv.org/abs/https://doi.org/10.1137/0125024} {https://doi.org/10.1137/0125024} \BibitemShut {NoStop}%
\bibitem [{\citenamefont {Wolf}(1974)}]{Wolf_1974}%
  \BibitemOpen
  \bibfield  {author} {\bibinfo {author} {\bibfnamefont {K.~B.}\ \bibnamefont {Wolf}},\ }\bibfield  {title} {\bibinfo {title} {Canonical transforms. i. complex linear transforms},\ }\href {https://doi.org/10.1063/1.1666811} {\bibfield  {journal} {\bibinfo  {journal} {Journal of Mathematical Physics}\ }\textbf {\bibinfo {volume} {15}},\ \bibinfo {pages} {1295} (\bibinfo {year} {1974})}\BibitemShut {NoStop}%
\bibitem [{\citenamefont {Rossmann}(2002)}]{Rossmann2002}%
  \BibitemOpen
  \bibfield  {author} {\bibinfo {author} {\bibfnamefont {W.}~\bibnamefont {Rossmann}},\ }\href {https://doi.org/10.1093/oso/9780198596837.001.0001} {\emph {\bibinfo {title} {Lie Groups: An Introduction Through Linear Groups}}}\ (\bibinfo  {publisher} {Oxford University Press},\ \bibinfo {year} {2002})\BibitemShut {NoStop}%
\bibitem [{\citenamefont {Wolf}(2004)}]{Wolf_2004}%
  \BibitemOpen
  \bibfield  {author} {\bibinfo {author} {\bibfnamefont {K.~B.}\ \bibnamefont {Wolf}},\ }\href@noop {} {\emph {\bibinfo {title} {Geometric optics on phase space}}}\ (\bibinfo  {publisher} {Springer Science \& Business Media},\ \bibinfo {year} {2004})\BibitemShut {NoStop}%
\bibitem [{\citenamefont {Hall}(2013)}]{Hall2013}%
  \BibitemOpen
  \bibfield  {author} {\bibinfo {author} {\bibfnamefont {B.~C.}\ \bibnamefont {Hall}},\ }\href {https://doi.org/10.1007/978-1-4614-7116-5} {\emph {\bibinfo {title} {Quantum Theory for Mathematicians}}}\ (\bibinfo  {publisher} {Springer New York},\ \bibinfo {year} {2013})\BibitemShut {NoStop}%
\bibitem [{\citenamefont {Hall}(2015)}]{Hall2015}%
  \BibitemOpen
  \bibfield  {author} {\bibinfo {author} {\bibfnamefont {B.~C.}\ \bibnamefont {Hall}},\ }\href {https://doi.org/10.1007/978-3-319-13467-3} {\emph {\bibinfo {title} {Lie Groups, Lie Algebras, and Representations: An Elementary Introduction}}}\ (\bibinfo  {publisher} {Springer International Publishing},\ \bibinfo {year} {2015})\BibitemShut {NoStop}%
\bibitem [{\citenamefont {Korneev}\ \emph {et~al.}(2025)\citenamefont {Korneev}, \citenamefont {Ramos-Prieto}, \citenamefont {Soto-Eguibar}, \citenamefont {Ruíz}, \citenamefont {Sánchez-de-la Llave},\ and\ \citenamefont {Moya-Cessa}}]{Korneev_2025}%
  \BibitemOpen
  \bibfield  {author} {\bibinfo {author} {\bibfnamefont {N.}~\bibnamefont {Korneev}}, \bibinfo {author} {\bibfnamefont {I.}~\bibnamefont {Ramos-Prieto}}, \bibinfo {author} {\bibfnamefont {F.}~\bibnamefont {Soto-Eguibar}}, \bibinfo {author} {\bibfnamefont {U.}~\bibnamefont {Ruíz}}, \bibinfo {author} {\bibfnamefont {D.}~\bibnamefont {Sánchez-de-la Llave}},\ and\ \bibinfo {author} {\bibfnamefont {H.~M.}\ \bibnamefont {Moya-Cessa}},\ }\bibfield  {title} {\bibinfo {title} {Unified approach to paraxial propagation in uniform media and media with linear or quadratic refractive index distribution},\ }\href {https://doi.org/10.1088/1402-4896/ade01d} {\bibfield  {journal} {\bibinfo  {journal} {Physica Scripta}\ }\textbf {\bibinfo {volume} {100}},\ \bibinfo {pages} {075515} (\bibinfo {year} {2025})}\BibitemShut {NoStop}%
\bibitem [{\citenamefont {Ramos-Prieto}\ \emph {et~al.}(2024)\citenamefont {Ramos-Prieto}, \citenamefont {de-la Llave}, \citenamefont {Ruíz}, \citenamefont {Arrizón}, \citenamefont {Soto-Eguibar},\ and\ \citenamefont {Moya-Cessa}}]{Ramos_GRIN_2024}%
  \BibitemOpen
  \bibfield  {author} {\bibinfo {author} {\bibfnamefont {I.}~\bibnamefont {Ramos-Prieto}}, \bibinfo {author} {\bibfnamefont {D.~S.}\ \bibnamefont {de-la Llave}}, \bibinfo {author} {\bibfnamefont {U.}~\bibnamefont {Ruíz}}, \bibinfo {author} {\bibfnamefont {V.}~\bibnamefont {Arrizón}}, \bibinfo {author} {\bibfnamefont {F.}~\bibnamefont {Soto-Eguibar}},\ and\ \bibinfo {author} {\bibfnamefont {H.}~\bibnamefont {Moya-Cessa}},\ }\bibfield  {title} {\bibinfo {title} {Cauchy–{R}iemann beams in {GRIN} media},\ }\href {https://doi.org/https://doi.org/10.1016/j.ijleo.2024.171864} {\bibfield  {journal} {\bibinfo  {journal} {Optik}\ }\textbf {\bibinfo {volume} {309}},\ \bibinfo {pages} {171864} (\bibinfo {year} {2024})}\BibitemShut {NoStop}%
\bibitem [{\citenamefont {Urz\'{u}a}\ \emph {et~al.}(2024)\citenamefont {Urz\'{u}a}, \citenamefont {Ramos-Prieto},\ and\ \citenamefont {Moya-Cessa}}]{Urzua_2024}%
  \BibitemOpen
  \bibfield  {author} {\bibinfo {author} {\bibfnamefont {A.~R.}\ \bibnamefont {Urz\'{u}a}}, \bibinfo {author} {\bibfnamefont {I.}~\bibnamefont {Ramos-Prieto}},\ and\ \bibinfo {author} {\bibfnamefont {H.~M.}\ \bibnamefont {Moya-Cessa}},\ }\bibfield  {title} {\bibinfo {title} {Integrated optical wave analyzer using the discrete fractional {F}ourier transform},\ }\href {https://doi.org/10.1364/JOSAB.533919} {\bibfield  {journal} {\bibinfo  {journal} {J. Opt. Soc. Am. B}\ }\textbf {\bibinfo {volume} {41}},\ \bibinfo {pages} {2358} (\bibinfo {year} {2024})}\BibitemShut {NoStop}%
\bibitem [{\citenamefont {Goncharenko}\ \emph {et~al.}(1991)\citenamefont {Goncharenko}, \citenamefont {Logvin}, \citenamefont {Samson}, \citenamefont {Shapovalov},\ and\ \citenamefont {Turovets}}]{Goncharenko}%
  \BibitemOpen
  \bibfield  {author} {\bibinfo {author} {\bibfnamefont {A.}~\bibnamefont {Goncharenko}}, \bibinfo {author} {\bibfnamefont {Y.}~\bibnamefont {Logvin}}, \bibinfo {author} {\bibfnamefont {A.}~\bibnamefont {Samson}}, \bibinfo {author} {\bibfnamefont {P.}~\bibnamefont {Shapovalov}},\ and\ \bibinfo {author} {\bibfnamefont {S.}~\bibnamefont {Turovets}},\ }\bibfield  {title} {\bibinfo {title} {Ermakov {H}amiltonian systems in nonlinear optics of elliptic {G}aussian beams},\ }\href {https://doi.org/https://doi.org/10.1016/0375-9601(91)90602-5} {\bibfield  {journal} {\bibinfo  {journal} {Physics Letters A}\ }\textbf {\bibinfo {volume} {160}},\ \bibinfo {pages} {138} (\bibinfo {year} {1991})}\BibitemShut {NoStop}%
\bibitem [{\citenamefont {Bertin}\ \emph {et~al.}(2012)\citenamefont {Bertin}, \citenamefont {Pimentel},\ and\ \citenamefont {Ramirez}}]{Bertin}%
  \BibitemOpen
  \bibfield  {author} {\bibinfo {author} {\bibfnamefont {M.~C.}\ \bibnamefont {Bertin}}, \bibinfo {author} {\bibfnamefont {B.~M.}\ \bibnamefont {Pimentel}},\ and\ \bibinfo {author} {\bibfnamefont {J.~A.}\ \bibnamefont {Ramirez}},\ }\bibfield  {title} {\bibinfo {title} {Construction of time-dependent dynamical invariants: A new approach},\ }\href {https://doi.org/10.1063/1.3702824} {\bibfield  {journal} {\bibinfo  {journal} {Journal of Mathematical Physics}\ }\textbf {\bibinfo {volume} {53}},\ \bibinfo {pages} {042104} (\bibinfo {year} {2012})}\BibitemShut {NoStop}%
\bibitem [{\citenamefont {Guasti}\ and\ \citenamefont {Moya-Cessa}(2003)}]{Guasti_2003}%
  \BibitemOpen
  \bibfield  {author} {\bibinfo {author} {\bibfnamefont {M.~F.}\ \bibnamefont {Guasti}}\ and\ \bibinfo {author} {\bibfnamefont {H.}~\bibnamefont {Moya-Cessa}},\ }\bibfield  {title} {\bibinfo {title} {Solution of the schrödinger equation for time-dependent 1d harmonic oscillators using the orthogonal functions invariant},\ }\href {https://doi.org/10.1088/0305-4470/36/8/305} {\bibfield  {journal} {\bibinfo  {journal} {Journal of Physics A: Mathematical and General}\ }\textbf {\bibinfo {volume} {36}},\ \bibinfo {pages} {2069} (\bibinfo {year} {2003})}\BibitemShut {NoStop}%
\bibitem [{\citenamefont {Gu\'ery-Odelin}\ \emph {et~al.}(2019)\citenamefont {Gu\'ery-Odelin}, \citenamefont {Ruschhaupt}, \citenamefont {Kiely}, \citenamefont {Torrontegui}, \citenamefont {Mart\'{\i}nez-Garaot},\ and\ \citenamefont {Muga}}]{Guery_2019}%
  \BibitemOpen
  \bibfield  {author} {\bibinfo {author} {\bibfnamefont {D.}~\bibnamefont {Gu\'ery-Odelin}}, \bibinfo {author} {\bibfnamefont {A.}~\bibnamefont {Ruschhaupt}}, \bibinfo {author} {\bibfnamefont {A.}~\bibnamefont {Kiely}}, \bibinfo {author} {\bibfnamefont {E.}~\bibnamefont {Torrontegui}}, \bibinfo {author} {\bibfnamefont {S.}~\bibnamefont {Mart\'{\i}nez-Garaot}},\ and\ \bibinfo {author} {\bibfnamefont {J.~G.}\ \bibnamefont {Muga}},\ }\bibfield  {title} {\bibinfo {title} {Shortcuts to adiabaticity: Concepts, methods, and applications},\ }\href {https://doi.org/10.1103/RevModPhys.91.045001} {\bibfield  {journal} {\bibinfo  {journal} {Rev. Mod. Phys.}\ }\textbf {\bibinfo {volume} {91}},\ \bibinfo {pages} {045001} (\bibinfo {year} {2019})}\BibitemShut {NoStop}%
\bibitem [{\citenamefont {Berry}(2009)}]{Berry_2009}%
  \BibitemOpen
  \bibfield  {author} {\bibinfo {author} {\bibfnamefont {M.~V.}\ \bibnamefont {Berry}},\ }\bibfield  {title} {\bibinfo {title} {Transitionless quantum driving},\ }\href {https://doi.org/10.1088/1751-8113/42/36/365303} {\bibfield  {journal} {\bibinfo  {journal} {Journal of Physics A: Mathematical and Theoretical}\ }\textbf {\bibinfo {volume} {42}},\ \bibinfo {pages} {365303} (\bibinfo {year} {2009})}\BibitemShut {NoStop}%
\bibitem [{\citenamefont {Mart\'{i}nez-Garaot}\ \emph {et~al.}(2017)\citenamefont {Mart\'{i}nez-Garaot}, \citenamefont {Muga},\ and\ \citenamefont {Tseng}}]{Martinez_2017}%
  \BibitemOpen
  \bibfield  {author} {\bibinfo {author} {\bibfnamefont {S.}~\bibnamefont {Mart\'{i}nez-Garaot}}, \bibinfo {author} {\bibfnamefont {J.~G.}\ \bibnamefont {Muga}},\ and\ \bibinfo {author} {\bibfnamefont {S.-Y.}\ \bibnamefont {Tseng}},\ }\bibfield  {title} {\bibinfo {title} {Shortcuts to adiabaticity in optical waveguides using fast quasiadiabatic dynamics},\ }\href {https://doi.org/10.1364/OE.25.000159} {\bibfield  {journal} {\bibinfo  {journal} {Opt. Express}\ }\textbf {\bibinfo {volume} {25}},\ \bibinfo {pages} {159} (\bibinfo {year} {2017})}\BibitemShut {NoStop}%
\bibitem [{\citenamefont {Sakurai}\ and\ \citenamefont {Napolitano}(2017)}]{Sakurai_2017}%
  \BibitemOpen
  \bibfield  {author} {\bibinfo {author} {\bibfnamefont {J.~J.}\ \bibnamefont {Sakurai}}\ and\ \bibinfo {author} {\bibfnamefont {J.}~\bibnamefont {Napolitano}},\ }\href {https://doi.org/10.1017/9781108499996} {\emph {\bibinfo {title} {Modern Quantum Mechanics}}}\ (\bibinfo  {publisher} {Cambridge University Press},\ \bibinfo {year} {2017})\BibitemShut {NoStop}%
\bibitem [{\citenamefont {Bagchi}\ and\ \citenamefont {Vinod}(2021)}]{Bagchi_2021}%
  \BibitemOpen
  \bibfield  {author} {\bibinfo {author} {\bibfnamefont {B.}~\bibnamefont {Bagchi}}\ and\ \bibinfo {author} {\bibfnamefont {A.}~\bibnamefont {Vinod}},\ }\bibfield  {title} {\bibinfo {title} {Ermakov-{P}inney equation for time-varying mass systems},\ }\href {https://doi.org/10.1088/1742-6596/2038/1/012002} {\bibfield  {journal} {\bibinfo  {journal} {Journal of Physics: Conference Series}\ }\textbf {\bibinfo {volume} {2038}},\ \bibinfo {pages} {012002} (\bibinfo {year} {2021})}\BibitemShut {NoStop}%
\bibitem [{\citenamefont {{Cruz y Cruz}}\ and\ \citenamefont {Gress}(2017)}]{Cruz_HGB_2017}%
  \BibitemOpen
  \bibfield  {author} {\bibinfo {author} {\bibfnamefont {S.}~\bibnamefont {{Cruz y Cruz}}}\ and\ \bibinfo {author} {\bibfnamefont {Z.}~\bibnamefont {Gress}},\ }\bibfield  {title} {\bibinfo {title} {Group approach to the paraxial propagation of {H}ermite–{G}aussian modes in a parabolic medium},\ }\href {https://doi.org/https://doi.org/10.1016/j.aop.2017.05.020} {\bibfield  {journal} {\bibinfo  {journal} {Annals of Physics}\ }\textbf {\bibinfo {volume} {383}},\ \bibinfo {pages} {257} (\bibinfo {year} {2017})}\BibitemShut {NoStop}%
\bibitem [{\citenamefont {Berry}(1984)}]{Berry_1984}%
  \BibitemOpen
  \bibfield  {author} {\bibinfo {author} {\bibfnamefont {M.~V.}\ \bibnamefont {Berry}},\ }\bibfield  {title} {\bibinfo {title} {Quantal phase factors accompanying adiabatic changes},\ }\href {https://doi.org/10.1098/rspa.1984.0023} {\bibfield  {journal} {\bibinfo  {journal} {Proceedings of the Royal Society of London. A. Mathematical and Physical Sciences}\ }\textbf {\bibinfo {volume} {392}},\ \bibinfo {pages} {45} (\bibinfo {year} {1984})}\BibitemShut {NoStop}%
\bibitem [{\citenamefont {Simon}\ and\ \citenamefont {Mukunda}(1993)}]{Simon_1993}%
  \BibitemOpen
  \bibfield  {author} {\bibinfo {author} {\bibfnamefont {R.}~\bibnamefont {Simon}}\ and\ \bibinfo {author} {\bibfnamefont {N.}~\bibnamefont {Mukunda}},\ }\bibfield  {title} {\bibinfo {title} {Bargmann invariant and the geometry of the g\"uoy effect},\ }\href {https://doi.org/10.1103/PhysRevLett.70.880} {\bibfield  {journal} {\bibinfo  {journal} {Phys. Rev. Lett.}\ }\textbf {\bibinfo {volume} {70}},\ \bibinfo {pages} {880} (\bibinfo {year} {1993})}\BibitemShut {NoStop}%
\bibitem [{\citenamefont {He}\ \emph {et~al.}(1995)\citenamefont {He}, \citenamefont {Friese}, \citenamefont {Heckenberg},\ and\ \citenamefont {Rubinsztein-Dunlop}}]{He_1995}%
  \BibitemOpen
  \bibfield  {author} {\bibinfo {author} {\bibfnamefont {H.}~\bibnamefont {He}}, \bibinfo {author} {\bibfnamefont {M.~E.~J.}\ \bibnamefont {Friese}}, \bibinfo {author} {\bibfnamefont {N.~R.}\ \bibnamefont {Heckenberg}},\ and\ \bibinfo {author} {\bibfnamefont {H.}~\bibnamefont {Rubinsztein-Dunlop}},\ }\bibfield  {title} {\bibinfo {title} {Direct observation of transfer of angular momentum to absorptive particles from a laser beam with a phase singularity},\ }\href {https://doi.org/10.1103/PhysRevLett.75.826} {\bibfield  {journal} {\bibinfo  {journal} {Phys. Rev. Lett.}\ }\textbf {\bibinfo {volume} {75}},\ \bibinfo {pages} {826} (\bibinfo {year} {1995})}\BibitemShut {NoStop}%
\bibitem [{\citenamefont {O'Neil}\ \emph {et~al.}(2002)\citenamefont {O'Neil}, \citenamefont {MacVicar}, \citenamefont {Allen},\ and\ \citenamefont {Padgett}}]{Oneil_2002}%
  \BibitemOpen
  \bibfield  {author} {\bibinfo {author} {\bibfnamefont {A.~T.}\ \bibnamefont {O'Neil}}, \bibinfo {author} {\bibfnamefont {I.}~\bibnamefont {MacVicar}}, \bibinfo {author} {\bibfnamefont {L.}~\bibnamefont {Allen}},\ and\ \bibinfo {author} {\bibfnamefont {M.~J.}\ \bibnamefont {Padgett}},\ }\bibfield  {title} {\bibinfo {title} {Intrinsic and extrinsic nature of the orbital angular momentum of a light beam},\ }\href {https://doi.org/10.1103/PhysRevLett.88.053601} {\bibfield  {journal} {\bibinfo  {journal} {Phys. Rev. Lett.}\ }\textbf {\bibinfo {volume} {88}},\ \bibinfo {pages} {053601} (\bibinfo {year} {2002})}\BibitemShut {NoStop}%
\bibitem [{\citenamefont {Krenn}\ \emph {et~al.}(2013)\citenamefont {Krenn}, \citenamefont {Fickler}, \citenamefont {Huber}, \citenamefont {Lapkiewicz}, \citenamefont {Plick}, \citenamefont {Ramelow},\ and\ \citenamefont {Zeilinger}}]{Krenn_2013}%
  \BibitemOpen
  \bibfield  {author} {\bibinfo {author} {\bibfnamefont {M.}~\bibnamefont {Krenn}}, \bibinfo {author} {\bibfnamefont {R.}~\bibnamefont {Fickler}}, \bibinfo {author} {\bibfnamefont {M.}~\bibnamefont {Huber}}, \bibinfo {author} {\bibfnamefont {R.}~\bibnamefont {Lapkiewicz}}, \bibinfo {author} {\bibfnamefont {W.}~\bibnamefont {Plick}}, \bibinfo {author} {\bibfnamefont {S.}~\bibnamefont {Ramelow}},\ and\ \bibinfo {author} {\bibfnamefont {A.}~\bibnamefont {Zeilinger}},\ }\bibfield  {title} {\bibinfo {title} {Entangled singularity patterns of photons in {I}nce-{G}auss modes},\ }\href {https://doi.org/10.1103/PhysRevA.87.012326} {\bibfield  {journal} {\bibinfo  {journal} {Phys. Rev. A}\ }\textbf {\bibinfo {volume} {87}},\ \bibinfo {pages} {012326} (\bibinfo {year} {2013})}\BibitemShut {NoStop}%
\bibitem [{\citenamefont {Plick}\ \emph {et~al.}(2013)\citenamefont {Plick}, \citenamefont {Krenn}, \citenamefont {Fickler}, \citenamefont {Ramelow},\ and\ \citenamefont {Zeilinger}}]{Plick_2013}%
  \BibitemOpen
  \bibfield  {author} {\bibinfo {author} {\bibfnamefont {W.~N.}\ \bibnamefont {Plick}}, \bibinfo {author} {\bibfnamefont {M.}~\bibnamefont {Krenn}}, \bibinfo {author} {\bibfnamefont {R.}~\bibnamefont {Fickler}}, \bibinfo {author} {\bibfnamefont {S.}~\bibnamefont {Ramelow}},\ and\ \bibinfo {author} {\bibfnamefont {A.}~\bibnamefont {Zeilinger}},\ }\bibfield  {title} {\bibinfo {title} {Quantum orbital angular momentum of elliptically symmetric light},\ }\href {https://doi.org/10.1103/PhysRevA.87.033806} {\bibfield  {journal} {\bibinfo  {journal} {Phys. Rev. A}\ }\textbf {\bibinfo {volume} {87}},\ \bibinfo {pages} {033806} (\bibinfo {year} {2013})}\BibitemShut {NoStop}%
\bibitem [{\citenamefont {Boyer}\ \emph {et~al.}(1975)\citenamefont {Boyer}, \citenamefont {Kalnins},\ and\ \citenamefont {Miller}}]{Boyer_1975}%
  \BibitemOpen
  \bibfield  {author} {\bibinfo {author} {\bibfnamefont {C.~P.}\ \bibnamefont {Boyer}}, \bibinfo {author} {\bibfnamefont {E.~G.}\ \bibnamefont {Kalnins}},\ and\ \bibinfo {author} {\bibfnamefont {J.}~\bibnamefont {Miller}, \bibfnamefont {W.}},\ }\bibfield  {title} {\bibinfo {title} {Lie theory and separation of variables. 7. the harmonic oscillator in elliptic coordinates and ince polynomials},\ }\href {https://doi.org/10.1063/1.522574} {\bibfield  {journal} {\bibinfo  {journal} {Journal of Mathematical Physics}\ }\textbf {\bibinfo {volume} {16}},\ \bibinfo {pages} {512} (\bibinfo {year} {1975})},\ \Eprint {https://arxiv.org/abs/https://pubs.aip.org/aip/jmp/article-pdf/16/3/512/19098483/512\_1\_online.pdf} {https://pubs.aip.org/aip/jmp/article-pdf/16/3/512/19098483/512\_1\_online.pdf} \BibitemShut {NoStop}%
\bibitem [{\citenamefont {Kong}\ \emph {et~al.}(2020)\citenamefont {Kong}, \citenamefont {Ying},\ and\ \citenamefont {Chen}}]{Kong_2020}%
  \BibitemOpen
  \bibfield  {author} {\bibinfo {author} {\bibfnamefont {Q.}~\bibnamefont {Kong}}, \bibinfo {author} {\bibfnamefont {H.}~\bibnamefont {Ying}},\ and\ \bibinfo {author} {\bibfnamefont {X.}~\bibnamefont {Chen}},\ }\bibfield  {title} {\bibinfo {title} {Shortcuts to adiabaticity for optical beam propagation in nonlinear gradient refractive-index media},\ }\bibfield  {journal} {\bibinfo  {journal} {Entropy}\ }\textbf {\bibinfo {volume} {22}},\ \href {https://doi.org/10.3390/e22060673} {10.3390/e22060673} (\bibinfo {year} {2020})\BibitemShut {NoStop}%
\bibitem [{\citenamefont {Li}\ \emph {et~al.}(2024)\citenamefont {Li}, \citenamefont {Paul}, \citenamefont {Novoa},\ and\ \citenamefont {Chen}}]{Li_2024}%
  \BibitemOpen
  \bibfield  {author} {\bibinfo {author} {\bibfnamefont {Y.}~\bibnamefont {Li}}, \bibinfo {author} {\bibfnamefont {K.}~\bibnamefont {Paul}}, \bibinfo {author} {\bibfnamefont {D.}~\bibnamefont {Novoa}},\ and\ \bibinfo {author} {\bibfnamefont {X.}~\bibnamefont {Chen}},\ }\bibfield  {title} {\bibinfo {title} {Shortcuts to adiabatic soliton compression in active nonlinear kerr media},\ }\href {https://doi.org/10.1364/OE.514457} {\bibfield  {journal} {\bibinfo  {journal} {Opt. Express}\ }\textbf {\bibinfo {volume} {32}},\ \bibinfo {pages} {7940} (\bibinfo {year} {2024})}\BibitemShut {NoStop}%
\end{thebibliography}
%

\end{document}